# Influence of twin structure on flux turbulence near the front of vortex annihilation.

L.S. Uspenskaya, I.G. Naumenko, A.A. Zhokhov

*Institute of Solid State Physics Russian Academy of Science, Chernogolovka, Moscow distr., 142432, Russia, uspenska@issp.ac.ru*

**Abstract**

The remagnetization of YBCO single crystals is studied by magneto-optical technique. Different behavior of the annihilation flux front in twinned and twin-less samples is observed. The specific induction distribution, called Meissner hole, is found to be necessary forerunner of turbulence development.

**Introduction**

Turbulence-like behavior of remagnetization flux front was first observed in 1994 [1] in *YBCO* single crystal. It looked like wavy distortion of a flat front between two opposite magnetized areas, which evolved with shape changing and "boiling" of magnetization near the flux front. Since that time this phenomenon was studied widely [2-7], and it was recognized general for type two superconductors. The dynamical instabilities of magnetic structures in anisotropic superconductors were predicted theoretically even before the first experimental observation. Theoretically the instabilities were expected to appear as a result of joint influence of an anisotropy and a non-linearity of the voltage-current characteristics [8] or due to a nonlinear dependence of resistivity upon an electric field [9]. Another scenario of turbulence development was suggested in Ref.[2,3]. Analyzing experimentally determined magnetic flux distribution the authors concluded that in the flux front region during the remagnetization the specific vortex-current configuration is formed: closed vortex loops surrounding an intensive current, which flows along the flux front, i.e. configuration similar to force free one [10]. The current was supposed to be identical to the Meissner current; and the whole configuration was called therefore Meissner hole (MH). This arrangement was assumed to be unstable with respect to transverse perturbations, i.e. with respect to a bending of the flux front line. So any defects and in particular, twin boundaries could cause the flux front distortion.

The importance of twins for turbulence formation was considered in Refs. [5,6,7,11]. It is known, that vortices move easier along than across twins [12,13,14]. Based on the existence of this twin-induced anisotropy, authors of Ref.[5,6,7] extended hydrodynamics of usual fluids on the motion of vortices and came to the conclusion that flux front turbulence could appear only in twined single crystal under the condition that twins are not parallel to the sample edges. Another consideration of vortex flow in frame of hydrodynamics leads to the conclusion that vortex-antivortex annihilation should be accompanied by thermal waves [15], which could cause a non-laminar vortex flow or turbulence too.

New observations of magnetic flux turbulence are reported in this paper. The study was performed on a set of YBCO single crystals with different twin structure varied from dense crossed twins to completely twin-less. Special attention was paid to the evolution of the induction distribution around the flux front and it's correlation with crystal structure. The turbulence like behavior of magnetization was found both in twin-less samples and in samples with regular unidirectional twins, and never in samples with dense crossed twins.

**Experimental technique**

The experiments were performed on high quality optimally doped *YBCO* single crystals with a thickness of 70 - *50 μm*, grown by the method of spontaneous crystallization from the melt [16]. The remagnetization of samples was studied by real time magneto-optic technique. The visualization of magnetic flux distribution was performed by means of yttrium-iron garnet film [17]. The pictures were recorded by video camera, which provides a time-resolution up to *0.04s*. The induction profiles were determined by direct measurements of an angle of Faraday rotation of polarized light [18], which was proportional to the induction value. So both the value and the direction of the induction vector were determined. The main experiments were performed on samples cooled under a field with the strength up to *3000 Oe*. In all experiments the field was directed perpendicular to the samples plane, which coincided with crystallographic *ab*-plane. The evolution of magnetic flux distribution with a time under heating after field cooling or remagnetization was studied.

**Turbulence development.**

The trapped magnetic flux distribution formed after field cooling is well known. After the magnetic field is switched off, the corresponding magnetic flux exits partially from the sample and an inverse magnetic flux enters the sample from the edges. The field gradient near the edges is determined by critical current value $J_c^{ab}$, while the depth of inverse magnetic flux penetration is determined by sample geometry [19]. If the sample has typical for single crystal *YBCO* shape of thin plate, then zero induction line, or the same, inverse magnetic flux front, is located on some distance from the edge inside the sample. The thinner is the sample the deeper an inverse magnetic flux enters the sample. Usually the inverse flux penetrates into a sample both under the virgin magnetization and under the remagnetization in the same manner in wide temperature range, i.e. the flux front has the same shape and the induction gradient is the same. However, in some crystals a remarkable difference between the flux distribution patterns in these two cases is observed. It is especially pronounced seen on samples, in which twin-less domains neighbors with twinned domains, Fig. 1.

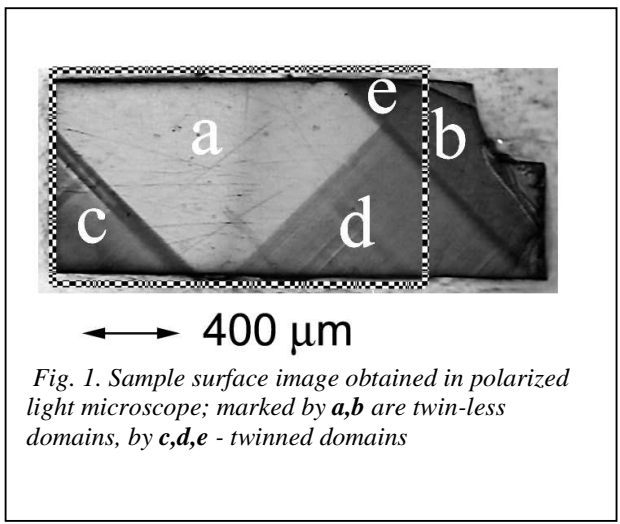

*Fig. 1. Sample surface image obtained in polarized light microscope; marked by **a,b** are twin-less domains, by **c,d,e** - twinned domains*

An example of virgin magnetization of this sample is shown in Fig. 2. The penetrating magnetic flux (bright area in the picture) has usual pillow-like shape with small distortions of the flux front near the boundaries between twinned and untwined domains. The distortion is a consequence of deeper penetration of the flux in twin-less sample areas compared with twinned areas, which is absolutely typical for *YBCO* single crystals at *T > 40 K* [12,13,14]. The lower the temperature the smaller this distortion of flux front. The induction gradient across the flux front in twinned and twin-less domains during virgin magnetization is also near the same. However at the remagnetization process a remarkable difference in flux penetration pattern comes into view.

A typical evolution of magnetic flux distribution during a remagnetization is shown in Figs. 3. Here slowly increasing magnetic field is applied to the sample cooled under the opposite directed field of 2500 Oe. The pictures are captured in polarized-light microscope with near crossed analyzer and polarizer. Therefore magnetic field of both directions makes the image lighter. So the higher is local magnetic induction, the brighter looks corresponding sample area in the pictures, and zero induction corresponds to dark area. Therefore flux front looks in the figure as a dark band. At the very beginning of remagnetization the flux front pattern is near the same in twinned and twin-less domains, compare marked by white arrows regions in Fig. 3a. With external field enhancement the difference in the patterns becomes visible. First, visible width of flux front turns into the band with remarkably variable width instead of having nearly the same width all along the front line, how it was observed at virgin magnetization. The band becomes wider in twin-less domain and remarkably narrower in twinned domains, compare flux front in points marked with white arrows in Figs. 3. Second, wavy distortions of flux front band appear in twinned domains while the flux front remains smooth even line in twin-less domain. And the more important point, the magnetic flux concentration all way along the flux front grows on both sides around the front in twinned domain. Black arrows in Figs. 3 mark some points in which the flux concentration could be seen easier. This flux concentration is a forerunner of the turbulence behavior of magnetic flux, which will be described below.

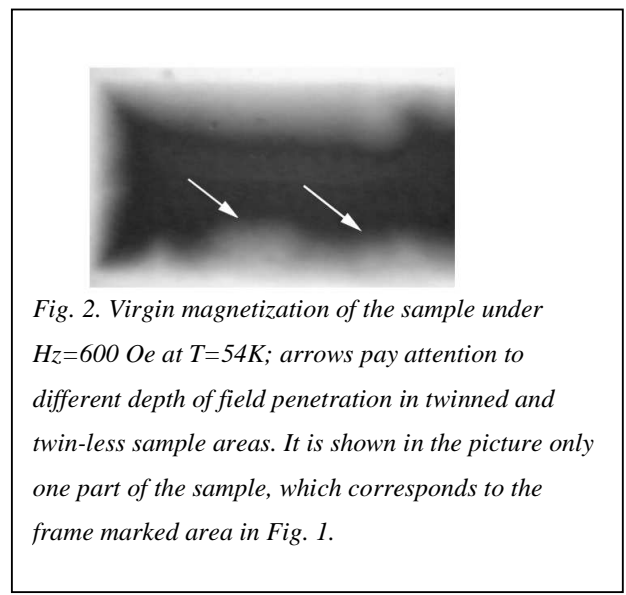

*Fig. 2. Virgin magnetization of the sample under $H_z=600$ Oe at T=54K; arrows pay attention to different depth of field penetration in twinned and twin-less sample areas. It is shown in the picture only one part of the sample, which corresponds to the frame marked area in Fig. 1.*

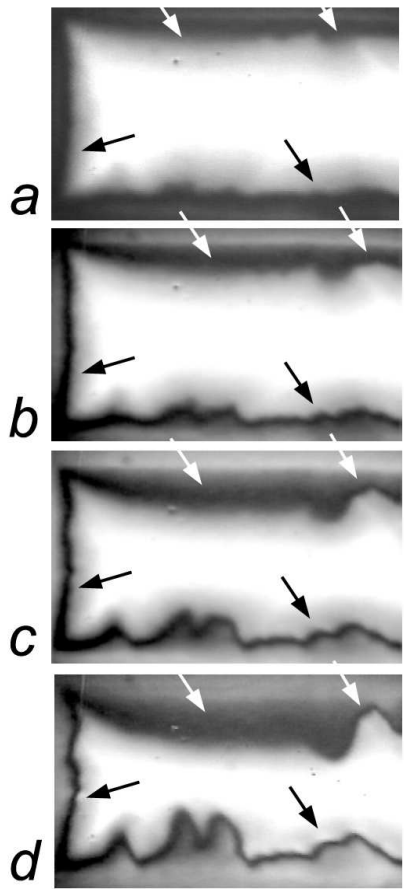

*Fig. 3. Remagnetization of the sample at T=54 K following 1200 Oe field cooling; $H_z$=215 Oe, 327 Oe, 450 Oe, 860 Oe correspondingly; white arrows point out the twinned and untwinned domain areas where the difference in flux front width is seen better; black arrows show some places where magnetic field is already concentrated near the flux front.*

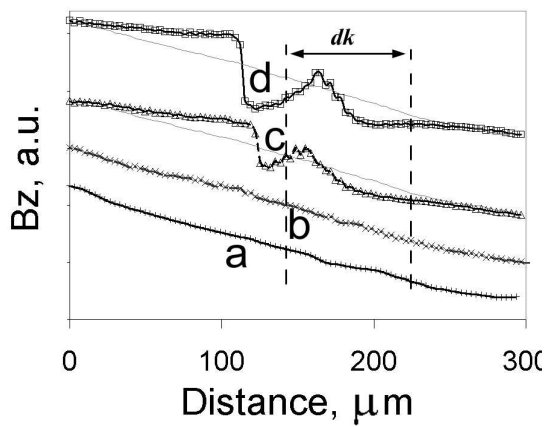

*Fig.4. Induction profiles $B_z(r)$ taken across entered magnetic flux at T=65 K in untwinned (a,b) and twinned (d,c) sample areas.*

The difference in the induction distribution around the flux front is seen more pronounced on the induction profiles $B_z(r)$ measured in the direction across the flux front in twinned and twin-less domains, Fig. 4. The profiles were taken at $T = 65$ K. Profiles *a* and *b* were taken in twin-less domain during virgin magnetization and during the remagnetization *10 s* after magnetic field was applied. Profiles *c* and *d* were taken in twinned domain during the remagnetization, *c* - in first second after magnetic field was applied, *d* - a few seconds later. The curves are shifted along the y-axis on the figure to better distinguish them. Curves *a* and *b* are smooth near linear function of the distance. Curves *c* and *d* are nonlinear function of the distance with a kink near the flux front, which grows with time. The kink corresponds to magnetic flux concentration around zero induction line described above. The time of non-linearity evolution, *dt*, depends upon the temperature. The higher is the temperature, the faster is the whole process, e.g. *dt(T=54K) ~ 1 min, dt(T=65K) ~ 10 s*. The distance, *dk*, up to which the deviation from linearity is extended rises with temperature, *dk(T=54K) ~ 20 μm, dk(T=65K) ~ 100 μm.*. The $B_z(r)$ profiles taken everywhere in the sample always remain smooth linear function like *a* and *b*-curves at virgin magnetization. The cited values, temperature range of turbulence development, time scale, and distances are relevant to this particular sample; they are different for other studied samples. However the major features of turbulence development are always reproducible.

Qualitatively the same evolution of flux front in twinned and untwinned areas is observed in both cases when the sample is slowly remagnetized by external field and when the magnetic flux trapped after field cooling relaxes slowly under heating: in the twin-less domain the flux front remains smooth band while in twinned domains narrow front is developed with wavy distortions of front line and flux concentrations near it.

Described evolution of flux front observed in twinned domains is typical for all samples with regular unidirectional twins or rarely crossed twins in the contrary to the samples with complicated twin structure, i.e. dense crossed twins.

**Meissner hole relaxation and motion**

Now it will be described the relaxation of the specific narrow flux front formed under slow remagnetization in twinned domains. Such front called Meissner hole (MH) was first described in [1]. It was found, that MH is not formed in untwined area during slow remagnetization. But MH can drift in the twin-less region from the twinned one along zero induction line. One example of such drift is shown in Fig. 5. First the sample was cooled under *+3000 Oe* field down to *T = 50 K*, then the field was slowly switched to *-200 Oe* and heating with 1 grad/min was started. The images of the same sample area, which include the flux front with MH and without MH, were taken with *10 s* interval one after another. Already during the inverse field application the MH was formed in twinned sample area. One end of MH, marked by white arrow in Fig. 5a, was attached exactly to the boundary between twinned domain (top part of images in Figs. 5) and twin-less domain (middle and bottom part of images in Figs. 5). The position of the flux front is nearly fixed because external field is kept constant while temperature is changed less than for *1 grad*. However, the end of MH drifts along the flux front in the same time interval for *~400 μm* from twinned domain into untwinned one. Naturally, at higher temperatures it moves faster. The propagation of MH into untwinned domain causes a magnetic flux concentration around the flux front.

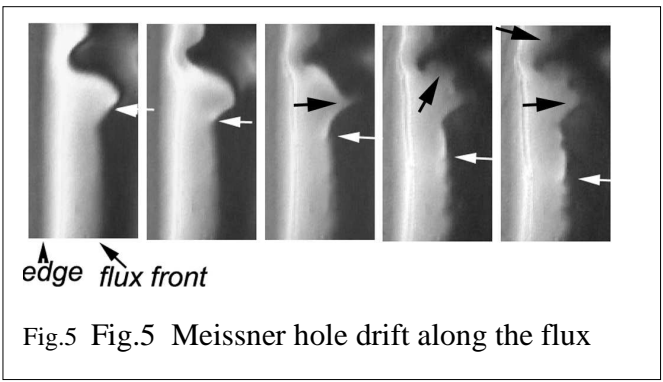

Fig.5  Fig.5  Meissner hole drift along the flux

The farther evolution of flux front structure goes in the same way in twinned and twin-less domains. The flux concentration, or the same vortices concentration, grows around the front, which carry MH, and simultaneously wavy distortions of flux front grows up. In some "weak" points, preferably in the bending of flux front, the vortices are concentrated more than in other places. At some moment they quickly coagulate in such points into bundle, which are seen near the flux front as small bright bubbles, e.g. see marked by black arrow points in Fig. 3d. The bundles carry magnetic flux with opposite direction, which corresponds to the induction vector directions on two sides of flux front. The bundles formation is accompanied by additional flux front bending. When the bundles grow large enough, or they concentrate large enough local magnetic field, they break locally the flux front line and annihilate. As a consequence, a discontinuity of MH line appears. It becomes subdivided into parts by patches with a smooth flux front. This process can be seen in top parts of images in Figs. 5, where the evolution from flux front with MH to smooth flux front takes place. The relaxed segments of front are marked in Figs. 5 by black arrows. The process of MH breaking is very fast, faster than *0.1 s*. Therefore it looks like "boiling" of magnetization around the flux line. Usually this process is called the turbulence of magnetic flux. Interesting, that MH does not rebuild in the area any more till new remagnetization. But the process proceeds near the flux front with residual fragments of MH. For a while, due the same mechanism of vortex concentration, attraction and breaking of residual segments of MH, all flux front becomes smooth band with linear variation of induction across it. Naturally, all process from the very beginning of vortex concentration to the complete relaxation takes different time in depend upon the temperature: the higher the temperature the faster the process. By the order of magnitude it takes about *~ 1 min* at $T = 50 K$ and *~ 10 s* at $T = 65 K$. So the turbulence can develop in untwinned area as well in twinned one even at slow remagnetization.

It was found also that the flux front carrying MH can propagate under an angle to twin boundaries, but cannot move perpendicular to them. For example, in the situation when twinned domain is located at some distance from the sample edge similar to *b* domain in Fig. 1, the magnetic flux enters the sample freely till the twin boundary. Father flux propagation depends upon flux front structure. The flux spreads deeper through twin boundary to the sample interior while virgin magnetization, but it is delayed by the boundary if it carry MH. In the last case vortices are concentrated on one side of MH only, namely on the side that is free from the twins. Then vortices coagulate into the bundles on that side of flux front. The bundles drift along the flux front to the sample edge and exit from the sample, reducing so the induction concentration near the flux front. No wavy front distortion is there. The flux front remains attached to flat boundary between twinned and twin-less regions. However the bundles are formed and drift fast along the flux front. This bundles drift is also a reflex of turbulence like behavior of magnetization in superconductors.

Pulse remagnetization as well as fast heating after field cooling cause turbulence development in wider temperature range, than slow remagnetization. For this particular sample pulse remagnetization expands the temperature range of turbulence development upto *40 K – 80 K* from *55 K – 65 K* at slow process. The specific flux front, called MH, is formed under pulse filed in shorter time than under slow field sweep, while the turbulence evolution takes the same time at the same temperature. Again, the turbulence is observed only in twin-less samples and in samples with rare parallel twins. No turbulence features was found in single crystals with complicated twin structure, i.e. with dense crossed twins. In details the turbulence development under pulse remagnetization and under fast heating will be described in the next publication.

**Discussion**

Found in the study conditions for turbulence development can be summarized as follows. The turbulence is observed in good quality single crystals with simple twin structure or without twins, on the boundary between opposite magnetized areas, but it is easier formed in samples with regular unidirectional twins. The whole process is observed at elevated temperatures, T>40 K. The temperature range of turbulence is narrow for slow remagnetization and wider for fast remagnetization. The specific induction distribution near the flux front, called Meissner hole, precedes the turbulence development. The higher is the temperature, the faster the specific non-linear induction distribution is formed around the flux front, and the faster the system evolves from that non-linearity through the turbulence to normal state. Under pulse magnetization the nonlinear induction distribution is formed faster than under slow remagnetization, but farther evolution to equilibrium takes the same time.

First requirement of good sample quality means that vortices should behave similar to continues lines that are difficult to disrupt. Really, the turbulence is not observed in BSCCO single crystals, in which vortices behave like weakly bounded pancakes. There is no turbulence in single-domain melt-textured YBCO, in which vortices are easily cutting due to large amount or large-scale non-superconducting inclusions. No turbulence is observed in single crystal YBCO with dense crossed twins, in which vortex cutting is also promoted.

Second terms, the turbulence is observed only at elevated temperatures around flux front that subdivides opposite magnetized areas. The vortices drift in these areas toward zero induction plane under the induction gradient and annihilate at this plane. There is no turbulence at low temperatures. The higher is the temperature the more intensive becomes this drift. It happens probably, that at some temperature the annihilation process becomes too slow compared with time that is necessary to all drifted vortices annihilate. So, vortex concentration at zero induction line begins. It follows from the experiment that such concentration begins easier in samples with regular twins than in twin-less samples. The twins contribute to the pinning. At *T>40 K,* the twin planes becomes important pinning centers which restrict the vortex motion across the twins, while there is no restriction for motion along twins [13]. Our experiment has shown the magnetic flux penetrates deeper in twinned sample area than in untwinned. It means that the vortex motion along the twin planes is even promoted. So the pinning in twinned sample becomes remarkably asymmetric. It was shown in Ref. [5-7] that this asymmetry cause to discontinuity of tangent component of vortex velocity, that could lead to turbulent motion of vortices.

However we see that at zero induction line during the vortex concentration some stable vortex complex is formed. That is very important. We call it Meissner hole following Ref.[1-3], but we are not sure that this complex really has such current-vortex configuration. This MH "subdivides" vortices with opposite magnetization. It means, that there is no "friction" between vortices and antivortices. It is suppressed. Nevertheless the turbulence is developed. The stability of vortex complex formed on flux front was demonstrated in our experiments. It influence on turbulence development was also demonstrated. So it should be taken into account at theoretical consideration. Brandt [20] has shown that induction lines during remagnetization generally form close loops in the plate-like samples. Vortex lines should follow the induction lines. That means the vortices should be curved in the same manner as induction lines. If the loops are formed in some manner, then suggested current concentration inside these vortex loops looks quite natural, as this configuration is very similar to the stable force-free configuration, considered in different aspects in Refs. [2,3,10].

Our experiments have shown MH is formed easier in samples with simple twin structure. Probably twin planes prevent the longitudinal spreading of vortices, and help so vortices looping round and form MH, or discontinuity of tangent vortex velocity make difficult vortex annihilation and promote its concentration near the flux front. Despite the reason of vortex concentration, it is important that some stable vortex structure is formed. If it is frozen, it exist infinitely long time. It can be moved with flux front and can be destroyed only by full remagnetization of the sample. The flux front that carry MH interacts with twins in its own way: it cannot drift transversely to twin boundaries till MH is destroid. In twined areas the Meissner hole distortion frequently takes place in points of twin intersections, which was mentioned also by Vlasko-Vlasov et al. [2,3]. The observed interaction of MH with twin boundaries give us the possibility to suggest an alternative explanation of the results obtained in Ref. [7]. If the crystal has the unidirectional twins with one sample edge strictly parallel to twin boundaries then MH would not be formed on that remagnetization flux front which is parallel to this edge, while it easily would be formed at other flux fronts. Therefore no turbulence development would be expected at this front, but at other flux fronts. So the observations [7] could be considered to be in agreement with our and [2,3]. The absence of the turbulence in crystals with dense crossed twins can be understood also in frame of MH influence on the turbulence development. MH would not be formed in such crystals, as vortices are cutting and reconnecting there very easy.

The last term looks obvious: the turbulence is developed in wider temperature range and in shorter time at fast remagnetization. It is clear that under the fast drop of the field or fast temperature increase the vortex flow is more intensive than under slow variation. So vortex concentration on the flux front begins easier. It just underlines that excess flow of vortex to flux front is necessary for turbulence development.

Of couse, different reasons could cause turbulence development in single crystals. From our point of view, all experimental results concerning turbulence development in superconductors obtained till now, could be described in terms of MH and its instabilities. It does not mean, that there could not be found another type of turbulence originated from hydrodynamic type of instabilities or other, predicted by different authors. Really it would be interesting to understand what kind of turbulence should appear at this or that material and what are necessary sample parameters.

In conclusion, we have observed different behavior of the annihilation flux front in twinned and twin-less domains in different samples and in the same sample and have found that the Meissner hole formation is important factor for turbulence development of the flux annihilation front. We come to the conclusion that the crystal structure determines if the Meissner hole and turbulence would appear. The Meissner hole is formed easier in crystals with regular unidirected twins and does not appear in the samples with dense crossed twins. We have found the Meissner hole is a rather stable object that can move with the flux front and can propagate along the flux front bringing turbulent behavior with it. However the mechanism of Meissner hole formation, its structure, the necessary conditions of its formation and the role of flux-current creep in the process still remains under the question. The mechanism of MH relaxation, the reason for vortex bundle formation is also under the question.

The work was supported by Russian Government (contract 40.012.1.1.4356), Russian Foundation of Fundamental Research (project 02-02-17062) and INTAS (project 02-2282). We would like to thank A.L. Rakhmanov, L.M. Fisher, V.A. Yampolskii, V.K. Vlasko-Vlasov, M.I. Indenbom and V.V. Ryazanov for useful discussions, and Humbolt foundation for grant of the optical polarized-light microscope and video camera.